\documentclass[twocolumn,trackchanges]{aastex701}

\usepackage{amsmath}
\usepackage{amssymb}
\usepackage{graphicx}
\usepackage{hyperref}
\usepackage{textcomp}
\usepackage{float}
\usepackage{multirow}
\usepackage{textcomp}

\begin{document}
\title{A Quasar Pair Sample Compiled from DESI DR1}

\author[0000-0003-1188-9573]{Liang Jing}
\affiliation{School of Physics and Astronomy, Beijing Normal University, Beijing, 100875, China}
\affiliation{Institute for Frontier in Astronomy and Astrophysics, Beijing Normal University, Beijing, 102206, China}
\email{202331160009@mail.bnu.edu.cn}

\author[0009-0006-9345-9639]{Qihang Chen}
\affiliation{School of Physics and Astronomy, Beijing Normal University, Beijing, 100875, China}
\affiliation{Institute for Frontier in Astronomy and Astrophysics, Beijing Normal University, Beijing, 102206, China}
\email{202131160006@mail.bnu.edu.cn}

\author[0009-0008-9072-4024]{Xingyu Zhu}
\affiliation{School of Physics and Astronomy, Beijing Normal University, Beijing, 100875, China}
\affiliation{Institute for Frontier in Astronomy and Astrophysics, Beijing Normal University, Beijing, 102206, China}
\email{202321160028@mail.bnu.edu.cn}

\author[0009-0008-8080-3124]{Zhuojun Deng}
\affiliation{School of Physics and Astronomy, Beijing Normal University, Beijing, 100875, China}
\affiliation{Institute for Frontier in Astronomy and Astrophysics, Beijing Normal University, Beijing, 102206, China}
\email{202431101076@mail.bnu.edu.cn}

\author[0000-0002-6684-3997]{Hu Zou}
\affiliation{Key Laboratory of Optical Astronomy, National Astronomical Observatories, Chinese Academy of Sciences, Beijing 100012, China}
\email{}

\author[0000-0003-4532-9523]{Jun-Qing Xia}
\affiliation{School of Physics and Astronomy, Beijing Normal University, Beijing, 100875, China}
\affiliation{Institute for Frontier in Astronomy and Astrophysics, Beijing Normal University, Beijing, 102206, China}
\email{}   

\author[0000-0002-8709-6759]{Jianghua Wu$^{\dagger}$}
\affiliation{School of Physics and Astronomy, Beijing Normal University, Beijing, 100875, China}
\affiliation{Institute for Frontier in Astronomy and Astrophysics, Beijing Normal University, Beijing, 102206, China}
\email[show]                                                                            {jhwu@bnu.edu.cn}

\begin{abstract} 

Interacting quasar pairs (QPs, either dual or binary) provide crucial insights into galaxy mergers, black hole growth, and large-scale structure formation. The current literature reports fewer than 200 spectroscopically confirmed QPs, highlighting the need for larger samples to enable statistically meaningful investigations. In this paper, we present a sample of 1,220 quasar pairs or candidates compiled from the DESI DR1 quasar sample. Among them, 145 systems have been previously reported. We visually classified the full sample using DESI Legacy Images and SPARCL spectra into three categories: QP (quasar pairs, N = 1020), QPC (quasar pair candidates, N = 142), LQC (lensed quasar candidates, N = 58). Within the LQC subset, we find an intriguing wide-separation ($\sim$7.15$^{\prime\prime}$) quadruply lensed quasar candidate. The redshift distribution of pair sample peaks at $z \sim 1$--$2.5$, with an overall pair fraction of $6.2^{+0.2}_{-0.2}\times10^{-4}$ (Poisson error) and a generally weak redshift dependence. 63.8\% of QPs have $|\Delta V_r| < 600$\,km/s, suggesting dynamical associations. This sample offers a statistically meaningful dataset for future studies of quasar pairs, lensing events, and potentially merger-triggered or merger-induced SMBH growth across the merger sequence.

\end{abstract}



\section{Introduction}

Galaxy mergers are transformative events that fundamentally reshape the structure, gas distribution, star formation, and overall evolutionary trajectories of galaxies \citep{1991ApJ...370L..65B,1996ApJ...471..115B,1994ApJ...431L...9M, 1996ApJ...464..641M,2005ApJ...630..705H,2008MNRAS.384..386C,2015MNRAS.447.2123C,2016ApJ...820...43S,2018MNRAS.478.3056B}. Although there is yet to be an observational consensus \citep{2009ApJ...691..705G,2011ApJ...726...57C,2012ApJ...744..148K,2012ApJ...751...72D,2013ApJ...779..136B,2014A&A...569A..37M,2016ApJ...830..156M,2018MNRAS.481..341S,2019MNRAS.487.2491E,2025arXiv251014743P}, a significant number of theoretical and observational works suggest mergers can and do trigger AGNs \citep{2007AJ....134..527W,2011MNRAS.418.2043E,2013MNRAS.435.3627E,2015MNRAS.451L..35E,2025OJAp....8E..12E,2014ApJ...795...62K,2014MNRAS.441.1297S,2017MNRAS.464.3882W,2023MNRAS.522.1736P,2025ApJ...989...73O}.
As these colossal systems interact, gravitational forces instigate intense bursts of star formation and drive gas and dust toward their central regions, thereby fueling the activity of supermassive black holes \citep{1991ApJ...370L..65B,1994ApJ...431L...9M,1996ApJ...464..641M,2008MNRAS.384..386C,2017MNRAS.469.4437C,2018MNRAS.478.3056B}. 
This process leads to the emergence of quasar pairs, defined as systems of two luminous AGNs, with transverse distances spanning from sub-pc to $\sim$110 kpc \citep{2025ApJS..281...25P}. These encompass dual AGNs (transverse distances of $\sim$0.03 -- 110 kpc), where both supermassive black holes (SMBHs) are simultaneously accreting but not necessarily gravitationally bound, and binary AGNs ($\lesssim$30 pc), in which the two SMBHs form a gravitationally bound Keplerian system \citep{2019NewAR..8601525D,2025ApJS..281...25P}. The eventual coalescence of such pairs is expected to generate detectable gravitational waves \citep{2017MNRAS.464.3131K,2023ApJ...952L..37A}.
Simulations further predict that during specific merger stages, dual or multiple quasar systems—where two or more SMBHs are simultaneously accreting—can emerge and remain active over timescales of tens to hundreds of millions of years \citep{2012ApJ...748L...7V,2023MNRAS.522.1895C,2025ApJS..281...25P}.
Compared with mergers hosting only a single AGN, dual AGNs show distinct growth patterns, with lower $M_{\rm BH}/M_\star$ ratios, higher specific star formation rates, and higher Eddington ratios \citep{2023MNRAS.522.1895C}.
Simulations also show that the number of dual AGNs cannot be explained purely by stochastic, secular processes, indicating that dual AGNs are key tracers of merger-driven SMBH growth \citep{2024Univ...10..237F}.  

A substantial number of articles have relied upon spectroscopic samples of AGNs/quasars for selection or confirmation of quasar pair systems \citep{2006AJ....131....1H,2010ApJ...719.1672H,2006ApJ...638..622M,2007ApJ...658...85M,2007ApJ...658...99M,2010ApJ...708..427L,2011ApJ...737..101L,2013ApJ...762..110L,2011ApJ...735...48S,Shen.2023,2012ApJ...746L..22K,2015ApJ...806..219C,2017ApJ...848..126S,2018MNRAS.480.1639D,2023MNRAS.519.5149D,2019ApJ...875..117P,2019ApJ...882...41H,2020ApJ...900...79H,2020ApJ...888...73H,2023ApJ...951...92B,2023ApJS..264....4M,2023ApJS..269...61D,2024MNRAS.531L..76Z,2025ApJS..277...49Z,2024A&A...692A.154W,2025arXiv250417777C}. Multi-wavelength observations of these systems have revealed complex interactions between merging galaxies and their central SMBHs, offering insights into both small-scale AGN triggering mechanisms and the broader context of large-scale structure formation \citep{2015ApJ...813..103M,2023MNRAS.519.5149D,2025arXiv250401251R}. 
Despite their importance, the systematic identification of dual and multiple quasar systems remains observationally challenging. A major source of contamination arises from stars, which can photometrically blend with quasars and appear as false positives in quasar catalogs \citep{2012ApJS..199....3R,2012AJ....144...49W,2020RNAAS...4..179Y,2024ApJS..271...54F}. The total number of confirmed quasar pairs is still relatively small, only $\sim$160 quasar pairs with a transverse distance of less than 110 kpc are confirmed up to 2020 \citep{2025ApJS..281...25P}.                               
Despite this contamination risk, several tens of genuine pairs have been identified serendipitously as by-products of lensed quasar surveys (e.g., \citealp{2003MNRAS.341....1M,2003MNRAS.341...13B,2006AJ....132..999O,2008AJ....135..496I,2012AJ....143..119I,2008ApJ...682..964B,2012MNRAS.419.2014J,2017MNRAS.472.5023L,2016MNRAS.456.1595M,2018MNRAS.475.2086A,2018MNRAS.480.1163S,2020A&A...636A..87C,2020MNRAS.494.3491L,2023MNRAS.520.3305L,2022MNRAS.510..500G,2022A&A...666A...1S,2023AJ....165..191Y,2023ApJS..269...61D,2024MNRAS.527.6253C,2025A&A...695A..76H}).
The existing samples are often heterogeneous, being constructed from serendipitous discoveries, high-resolution imaging, or spatially resolved spectroscopy, and are typically biased toward specific merger stages or AGN luminosity regimes \citep{2025ApJS..281...25P}. This will hinder robust statistical analyses and obscure the full diversity of quasar pair systems.
To overcome these obstacles, it is imperative to pursue deeper, more homogeneous surveys and refine detection techniques, which will ultimately improve models of galaxy evolution and black hole growth.

Recently, the Dark Energy Spectroscopic Instrument (DESI) released its first data release (DR1), containing 1.6 million spectroscopically confirmed quasars \citep{2025arXiv250314745D}. With its large sky coverage, high spectral quality, and systematic target selection, DR1 offers a deep and statistically uniform quasar sample well suited for population studies. In this work, we perform a self-matching analysis of the DR1 quasar catalog to search for quasar pairs and lensed quasars systematically.   

The paper is organized as follows. The spectroscopically confirmed quasar catalog exploited in this work and the method details are introduced in Sec.\ref{sec2}, which also describes our visual inspection and classification scheme and provides representative examples. Sec.~\ref{sec3} analyzes the basic statistical properties and presents a promising wide-separation quadruply lensed quasar candidate. We summarize this work in Sec.\ref{sec4}. A flat $\Lambda$CDM cosmology with $\Omega_{\Lambda}$ = 0.7, $\Omega_M$ = 0.3, and \textsl{H$_0$} = 70 km s$^{-1}$ Mpc$^{-1}$ is adopted throughout this paper.

\section{Data and Method} \label{sec2}
\setcounter{footnote}{0}
\subsection{DESI DR1 Spectroscopic Parent Sample} \label{sec2.1}

The Dark Energy Spectroscopic Instrument (DESI) is a state-of-the-art multi-object spectrograph mounted on the Mayall 4-meter telescope at Kitt Peak National Observatory (KPNO) \citep{2013arXiv1308.0847L,2016arXiv161100036D,2019AJ....157..168D,2022AJ....164..207D}. Its first major data release, DESI DR1, marks a significant milestone in spectroscopic surveys by providing high-quality spectra for millions of celestial objects, including 13.1M galaxies, 4M stars, and 1.6M quasars \citep{2025arXiv250314745D}. In this work, we utilize the high-quality quasar spectra provided in this release, which comprises quasars out to z $\approx 4$ for 1,650,736 sources and 2,182,309 spectra\footnote{\url{https://data.desi.lbl.gov/public/dr1/survey/catalogs/dr1/QSO/iron/}}, offering an unparalleled resource for astrophysical research.

\subsection{Candidate Quasar Pair Matching} \label{sec2.2}

In this study, we matched the catalog onto itself and then imposed the transverse distance and radial velocity difference criteria described below to search for quasar pairs from the DESI DR1 quasar catalog.  
To ensure physical association, we calculated for each quasar a redshift-dependent threshold of angular distance corresponding to a maximum transverse distance of 110\,kpc \citep{2025ApJS..281...25P}, formalized as:
\begin{equation}
    r_p \lesssim 110\, \mathrm{kpc},
    \label{eq:110kpc}
\end{equation}
where $r_p$ denotes the transverse distance between pair components. This threshold defines a dynamical search radius (this radius is based on the results from Illustris-TNG100 interacting pairs of galaxies from \citealp{2024MNRAS.529.1493P} and was defined in \citealp{2025ApJS..281...25P}) within which we search for neighboring quasars in the same catalog, i.e., perform self-matching to identify candidate pairs.

Next, we impose a radial velocity difference threshold to distinguish physically associated quasars from projected cases (\citealp{2006AJ....131....1H,2010ApJ...719.1672H}; see also \citealp{2025ApJS..281...25P}). Specifically, we require
\begin{equation}
|\Delta V_r| \lesssim 2000\,\mathrm{km\,s^{-1}},
\label{eq:2000kms}
\end{equation}
where $\Delta V_r$ is the line-of-sight radial velocity difference between pair components, computed following \citet{1999astro.ph..5116H}.
It accounts for peculiar velocities in dense environments, which can be as high as approximately $\sim500 \, \mathrm{km \, s^{-1}}$, and for broad-line region redshift uncertainties of up to $\sim1500 \, \mathrm{km \, s^{-1}}$ due to blueshifted emission lines \citep{2002AJ....124....1R}. This criterion is critical for distinguishing pairs that are likely at the same redshift from those that are merely chance alignments along the line of sight. 

After applying the $r_p$ and $|\Delta V_r|$ selection criteria \ref{eq:110kpc} and \ref{eq:2000kms}, we obtained a parent sample of 1,842 candidate quasar pair systems from the DESI DR1 quasar catalog, which is further classified in the following subsections.

\subsection{Visual Classification of Candidate Quasar Pairs}\label{sec2.3}
We perform a detailed visual inspection using imaging data from the ninth public data release of the DESI Legacy Imaging Surveys (LS-DR9; \citealp{2017PASP..129f4101Z,2019AJ....157..168D}) to classify the sample.
The SPARCL tool\footnote{\url{https://astrosparcl.datalab.noirlab.edu/}} contains spectral data from SDSS-DR16, SDSS-DR17, BOSS-DR16 \citep{2022ApJS..259...35A}, and DESI-EDR \citep{2023arXiv230606308D}. After matching with SPARCL using the x-match function in DataLab\footnote{\url{https://datalab.noirlab.edu/}} \citep{2014SPIE.9149E..1TF,2020A&C....3300411N}, we obtained a sparcl\_id, which can then be used in SPARCL programs to retrieve various selectable information about the target object, such as redshift, spectype, wavelength, flux, and much more \citep{2024arXiv240105576J}. These spectral data enable the creation of detailed spectral plots for quasars, providing critical insights into their properties and behaviors.

This visual verification is critical for assessing candidate quality. Those pairs with clearly resolved nuclei and distinct spectral features—such as different emission or absorption lines—are flagged as potential quasar pairs \citep{2025ApJS..281...25P}. In contrast, systems displaying lensing-like configurations (e.g., arcs, symmetric image) and highly similar spectra with nearly constant flux ratios are classified as lensed quasar candidates, as such configurations are consistent with gravitational lensing of a single background source (e.g., \citealp{2000PhRvD..62h4003V,2004PhDT.........1O,2017ApJ...851...48S,2020MNRAS.494.3491L,2022A&A...662A...4S}). To facilitate reliable classification, we examined the LS-DR9 model and residual images and plotted spectral flux ratios, all of which contribute to lens identification. In particular, the residual images help uncover faint or symmetric structures indicative of lensing, especially in compact systems \citep{2006ApJ...638..703B,2014ApJ...785..144G}. The full candidate selection and verification process is illustrated in Figure \ref{Flowchart}.

\begin{figure*}
    \centering
    \includegraphics[width=\textwidth]{Flowchart.png}
    \caption{Flowchart of the quasar pair sample selection process.}
    \label{Flowchart}
\end{figure*}

We then divided the visually examined quasar pair candidates into three subcategories based on their configurations and spectroscopic characteristics:

\begin{itemize}
    \item \textbf{QP (Quasar Pair):} Systems with angular separations $\geq 2^{\prime\prime}$ between pair components, where no prominent lensing galaxy is visible between or near the components, and the two quasars exhibit noticeably different spectral features. Taking into account the 1.5$^{\prime\prime}$ angular diameter of the DESI fiber and typical seeing conditions ($\sim 1–1.7^{\prime\prime}$; \citealp{2019AJ....157..168D,2025arXiv250314745D}), we adopt 2$^{\prime\prime}$ as a lower limit of the angular separation between pair components. It is worth noting that this relatively liberal choice allows for a larger sample size but opens up the possibility for fiber spillover and contaminants \citep{2018A&A...610L...7H,2020A&A...639A.117H,2023ApJ...945..167P,2023ApJ...954..116P}.

    \item \textbf{QPC (Quasar Pair Candidate):} Systems with angular separations $< 2^{\prime\prime}$, where the image clearly reveals two components without any apparent foreground lensing structure. Additionally, the flux ratio between the two spectra varies significantly with wavelength, suggesting two distinct sources. However, at these separations, the fiber spillover and contaminants are expected to be severe \citep{2020A&A...639A.117H,2023ApJ...945..167P}, so we treated such systems as quasar pair candidates. These systems will still likely require follow-up, spatially resolved spectroscopic observations for confirmation.
    
    \item \textbf{LQC (Lensed Quasar Candidate):} Systems over various angular separations exhibiting configurations or spectroscopic features indicative of gravitational lensing. These candidates are flagged as lensing candidates due to the lack of lens modeling or high-resolution image confirmation.

\end{itemize}

In addition, there are also some systems with very small angular separations ($\lesssim 1.2\arcsec$). Most of these systems are probably irrelevant contaminants either because they are observations of the same object, or because these systems exhibit very poor spectral quality (when two objects can be observed). We excluded these systems from subsequent analyses and provided brief notes on them in Appendix \ref{app:contaminant}.

\subsection{Examples}\label{sec2.4}
In this subsection, we present one representative system for each class of system described in Section \ref{sec2.3}. For each example, we show the DESI Legacy Survey images and the DESI spectra of the components, and we briefly summarize the key morphological and spectroscopic features that motivate its classification.
\subsubsection{Quasar Pair: J0145+0024} \label{sec2.4.1}
\begin{figure}[H]
    \centering
    \includegraphics[width=0.48\textwidth]{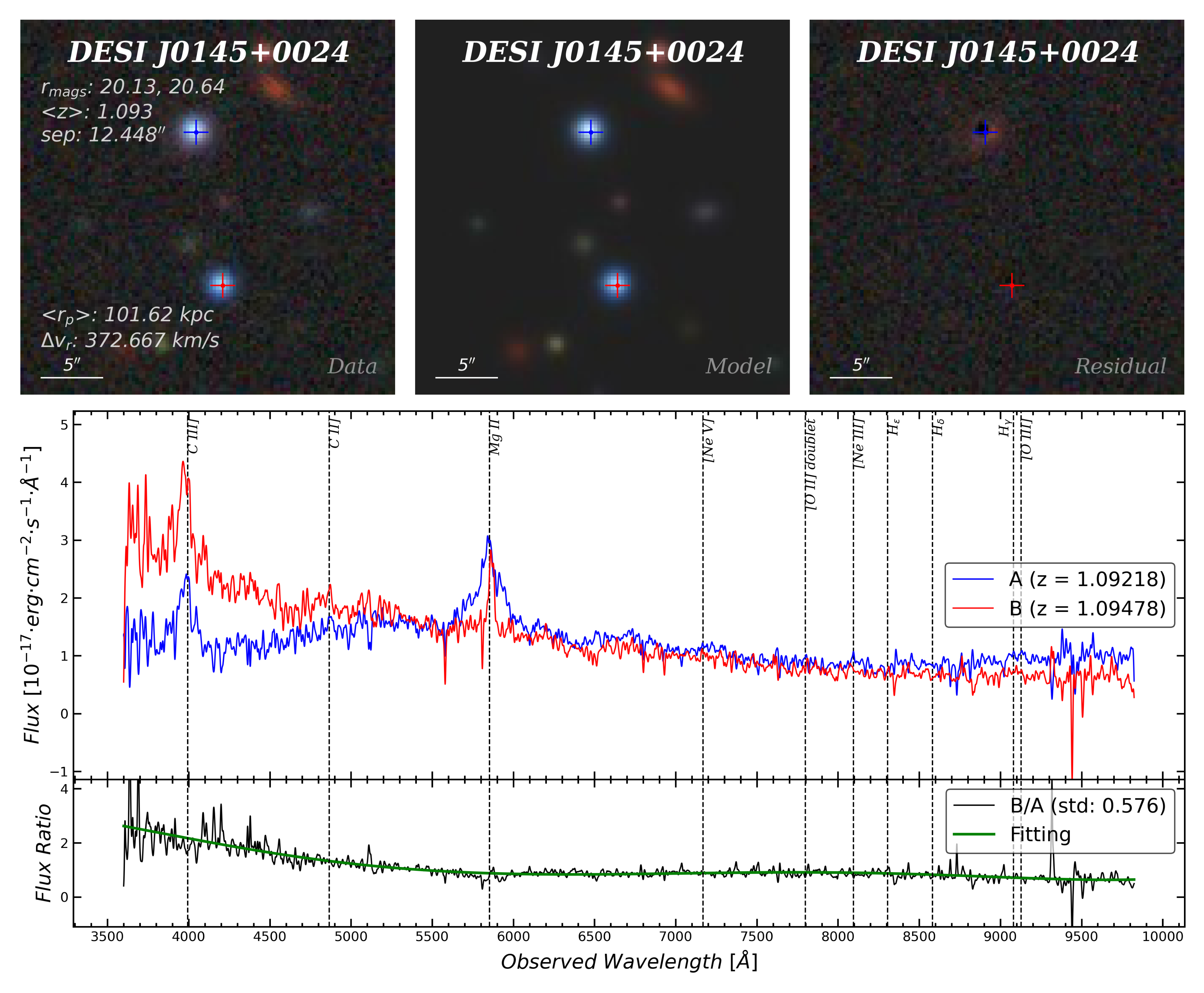}
    \caption{Example of a quasar pair, J0145+0024, discussed in Subsection \ref{sec2.4.1}.} The top-left panel shows the LS-DR9 image of J0145+0024, with the two sources marked by crosses (blue for source A and red for source B). The plot also annotates the $r$-band magnitudes ($r_{\mathrm{mag}}$) from the Legacy Survey, the average redshift ($\left<z\right>$), the angular separation ($\mathrm{sep}$), the average transverse distance ($\left<r_p\right>$), and the radial velocity difference ($\Delta V_r$). The top-middle and top-right panels display the LS-DR9 model and residual images, respectively, with the same cross markers indicating the source positions. The bottom panels present the spectra of the two quasars and their flux ratio. To better visualize variations in the flux ratio, we applied a polynomial fit and calculated the standard deviation (std) of the residuals, providing a measure of flux similarity.
    \label{ex_QP}
\end{figure}

J0145+0024, shown in Figure \ref{ex_QP}, exhibits two quasars at a relatively large angular separation. Their spectra are clearly distinct in continuum slope at short wavelengths and in width of Mg II emission. There are two objects lying between them, which, in principle, might form a lensing system with the two quasar images. The differential reddening or variability with time delay of the two images might account for the different slope at short wavelengths, but can hardly explain the similar continuum slope and strength longer than 5300 {\AA} and the difference in Mg II width. So we classify this system as a QP.

\subsubsection{Quasar Pair Candidate: J0118$-$0104} \label{sec2.4.2}
\begin{figure}[H]
    \centering
    \includegraphics[width=0.48\textwidth]{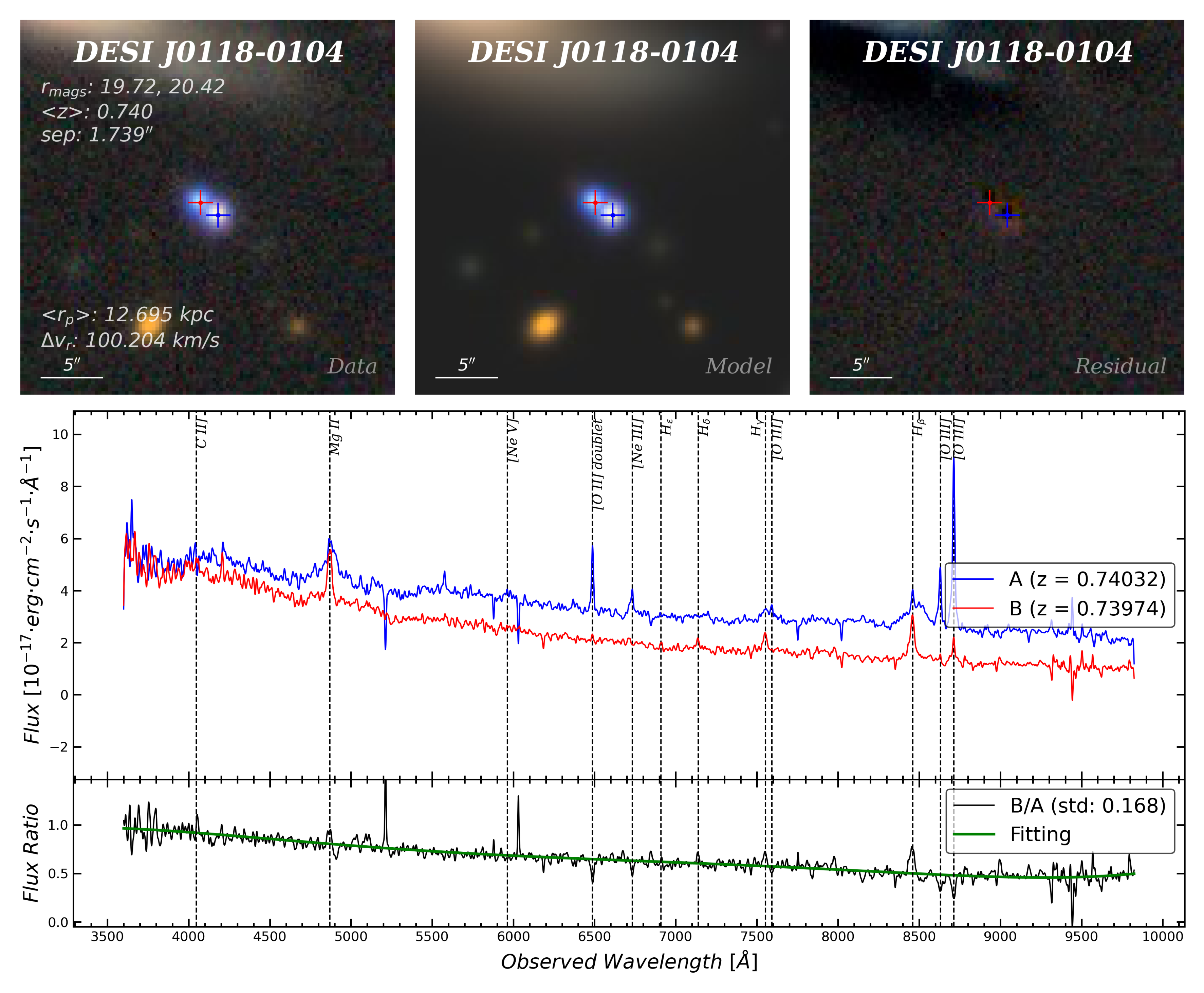}
    \caption{Example of a quasar-pair candidate, J0118$-$0104, discussed in Subsection~\ref{sec2.4.2}.
The image types of all panels are the same as in Figure~\ref{ex_QP}. 
There is no obvious intervening lensing structure between the two quasars, and the emission lines of source A are significantly broader than those of source B.
However, the components are separated by only 1.74$^{\prime\prime}$, comparable to the 1.5$^{\prime\prime}$ DESI fiber diameter, so the DESI spectra are likely affected by mutual fiber spillover \citep{2018A&A...610L...7H,2023ApJ...945..167P}. 
Given these caveats, we classify this system as QPC.}
    \label{ex_QPC}
\end{figure}

J0118$-$0104, shown in Figure \ref{ex_QPC}, exhibits two quasars with a small angular separation, close to the DESI fiber diameter (1.5$^{\prime\prime}$), which poses observational challenges due to potential fiber spillover effect \citep{2018A&A...610L...7H,2023ApJ...945..167P}. The LS-DR9 image shows two resolved components with no obvious massive foreground galaxy between or near them that could act as a lens. The two spectra have slight differences in continuum strength but marked differences in emission lines. The red spectrum have much narrower Mg II and H{$\beta$} lines than the blue spectrum, which may classify the fainter object as a type II quasar or narrow-line Seyfert 1 galaxy. The [O II] {$\lambda$}3727{\AA}, [Ne III] {$\lambda$}3869{\AA}, and H{$\beta$} {$\lambda$}4861{\AA} lines appear in the blue spectrum but are absent in the red one, giving further evidence that the two objects are distinct ones. Given the nearly same continuum slope of the two spectra, the differential reddening can hardly play a role in making the difference in the two spectra, especially the difference in the emission lines. A microlensing effect will produce simultaneous variability in both continuum and emission lines but not in the appearance and disappearance of the emission lines. The variability with time delay of two images and of continuum and emission lines may account for all differences between the two spectra, but the result should be a stronger continuum with weaker or absent emission lines (the stronger continuum outshines the emission lines due to the lagged response of the brightening of the emission lines to that of the continuum), which is opposite to the fact manifested in Figure \ref{ex_QPC}. As a result, we classify this system as a QPC with a high probability of being a genuine QP. It should be noted that this system was previously classified by \citet{2020MNRAS.494.3491L} as a nearly identical quasar (NIQ) system based on two spectra with relatively low SNR, which did not reveal the marked difference in the emission lines in Figure \ref{ex_QPC}.

\subsubsection{Lensed Quasar Candidate: J0941+0518} \label{sec2.4.3}
\begin{figure}[H]
    \centering
    \includegraphics[width=0.48\textwidth]{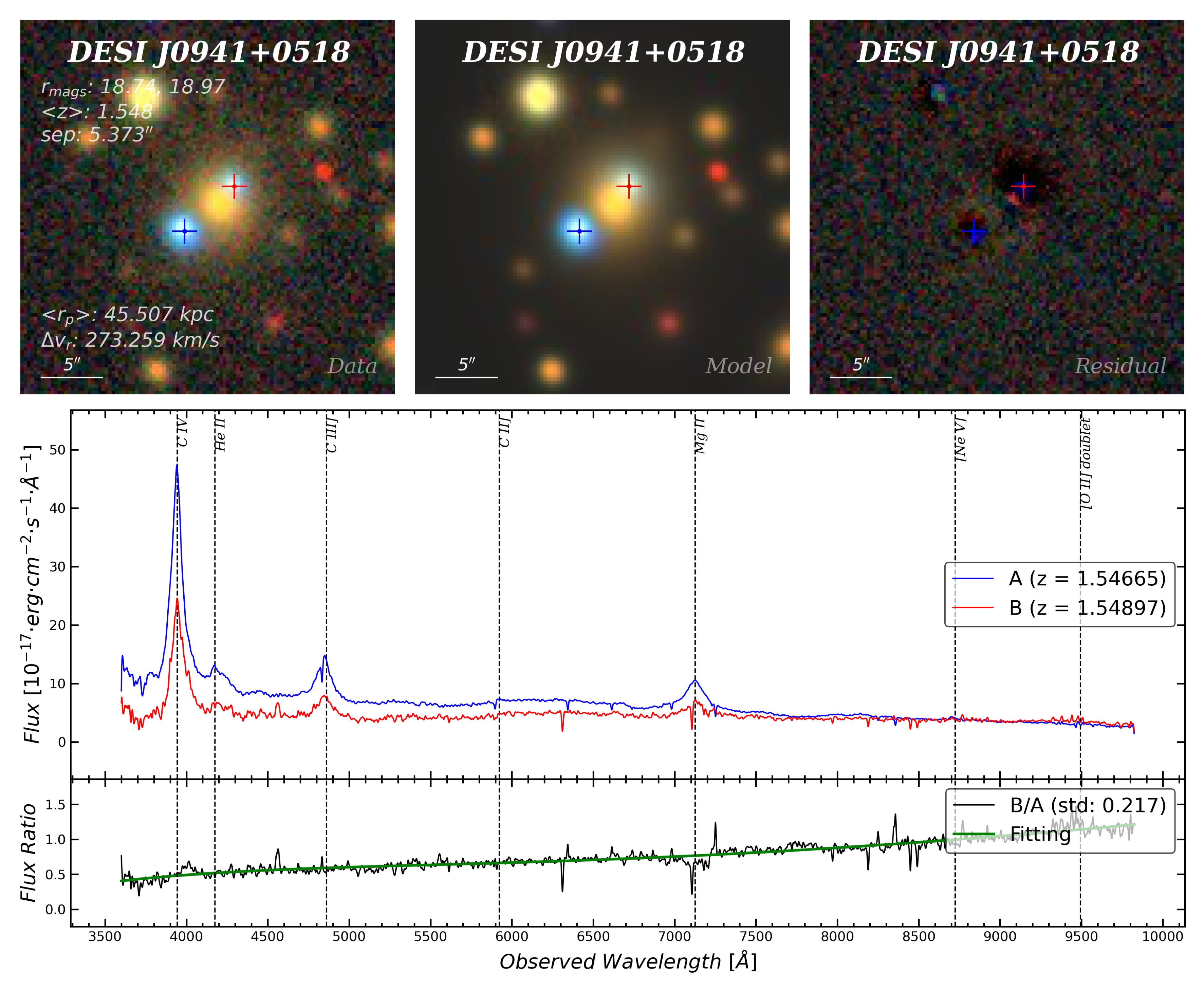}
    \caption{Example of a lensed quasar candidate, J0941+0518, discussed in Subsection \ref{sec2.4.3}.} The two quasars have nearly identical spectral features, suggesting they may be multiple images of the same background source. A bright, red galaxy is clearly visible between them, likely serving as the lensing object. Source B shows some absorption features, this may be due to its alignment with the foreground galaxy.
    \label{ex_LQC}
\end{figure}

J0941+0518, shown in Figure \ref{ex_LQC}, appears to have a massive, bright red galaxy situated between the two quasars, and the spectral features of the quasars are remarkably similar. In addition, the LS-DR9 residual image reveals a faint but coherent signature consistent with a lensing structure at the position of the intervening galaxy. The spectral flux ratio between the two quasars is also nearly constant across wavelengths, further supporting the lensing interpretation. Based on these combined morphological and spectroscopic indicators, we classify this system as an LQC. This system was previously reported by \citet{2018MNRAS.479.5060L} and confirmed as a gravitational lens, lending further support to our classification.

\subsection{Sample Overview}\label{sec2.5}
The final quasar pair sample comprises 1,220 systems that satisfy both the transverse distance and radial velocity criteria in Section \ref{sec2.2}, and is further classified into 1020 QPs, 142 QPCs and 58 LQCs based on visual inspection of the source morphology, fiber placement, and fiber separation. To identify previously studied systems within our sample, we first cross-matched our catalog with the Strong Gravitational Lens Database (SLED\footnote{\url{https://sled.amnh.org/}}) using a 1$^{\prime\prime}$ positional tolerance, and then matched the remaining sources against the Big Multi-AGN Catalog (Big MAC; \citealp{2025ApJS..281...25P}). We subsequently queried SIMBAD \citep{2000A&AS..143....9W} and the NASA/IPAC Extragalactic Database (NED\footnote{\url{https://ned.ipac.caltech.edu/}}) for the residual unmatched systems using the same 1$^{\prime\prime}$ radius. As a result, we identified 89 sources in SLED, 38 sources in the Big MAC, and 18 in NED, while SIMBAD did not yield any new associations. In total, this cross-matching yields 145 systems with previous identifications, including 23 confirmed lensed quasars and 122 candidate lensed quasars or quasar pairs. These matches are flagged in our catalog, and the corresponding classifications reported in the original references are included in a dedicated column for cross-reference. The column descriptions of the sample are summarized in Table \ref{fulltable}, and the full sample can be accessed at \url{https://github.com/astroliang/DESI_DR1_QP}. 

\begin{table*}[htbp] 
\scriptsize
\centering
\caption{Column descriptions of the quasar pair sample.}
\label{fulltable}
\begin{tabular}{p{0.22\textwidth}p{0.25\textwidth}|p{0.22\textwidth}p{0.25\textwidth}}
\hline
\textbf{Column} & \textbf{Description} & \textbf{Column} & \textbf{Description} \\
\hline\hline
System & System coordinate name & RA\_B\_sexa & Right Ascension of QSO B (sexagesimal) \\
SysCenterName & Complete system coordinate name & DEC\_B\_sexa & Declination of QSO B (sexagesimal) \\
TARGETID\_A & DESI target ID of QSO A & Z\_B & Redshift of QSO B \\
CoordUqName\_A & Coordinate name of QSO A & ZERR\_B & Redshift uncertainty of QSO B \\
RA\_A & Right Ascension of QSO A (deg) & ZWARN\_B & Redshift warning flag for QSO B \\
DEC\_A & Declination of QSO A (deg) & mag\_g\_B & g-band magnitude of QSO B \\
RA\_A\_sexa & Right Ascension of QSO A (sexagesimal) & mag\_r\_B & r-band magnitude of QSO B \\
DEC\_A\_sexa & Declination of QSO A (sexagesimal) & mag\_z\_B & z-band magnitude of QSO B \\
Z\_A & Redshift of QSO A & mag\_W1\_B & WISE W1-band magnitude of QSO B \\
ZERR\_A & Redshift uncertainty of QSO A & mag\_W2\_B & WISE W2-band magnitude of QSO B \\
ZWARN\_A & Redshift warning flag for QSO A & flux\_g\_B & g-band flux of QSO B \\
mag\_g\_A & g-band magnitude of QSO A & flux\_r\_B & r-band flux of QSO B \\
mag\_r\_A & r-band magnitude of QSO A & flux\_z\_B & z-band flux of QSO B \\
mag\_z\_A & z-band magnitude of QSO A & flux\_W1\_B & WISE W1-band flux of QSO B \\
mag\_W1\_A & WISE W1-band magnitude of QSO A & flux\_W2\_B & WISE W2-band flux of QSO B \\
mag\_W2\_A & WISE W2-band magnitude of QSO A & flux\_ivar\_g\_B & Inverse variance of g-band flux (B) \\
flux\_g\_A & g-band flux of QSO A & flux\_ivar\_r\_B & Inverse variance of r-band flux (B) \\
flux\_r\_A & r-band flux of QSO A & flux\_ivar\_z\_B & Inverse variance of z-band flux (B) \\
flux\_z\_A & z-band flux of QSO A & flux\_ivar\_W1\_B & Inverse variance of W1-band flux (B) \\
flux\_W1\_A & WISE W1-band flux of QSO A & flux\_ivar\_W2\_B & Inverse variance of W2-band flux (B) \\
flux\_W2\_A & WISE W2-band flux of QSO A & MORPHTYPE\_B & Morphological type of QSO B \\
flux\_ivar\_g\_A & Inverse variance of g-band flux (A) & SPECTYPE\_B & Spectral classification of QSO B \\
flux\_ivar\_r\_A & Inverse variance of r-band flux (A) & LASTNIGHT\_B & Observation date of QSO B \\
flux\_ivar\_z\_A & Inverse variance of z-band flux (A) & sep\_AB & Angular separation between QSO A and B \\
flux\_ivar\_W1\_A & Inverse variance of W1-band flux (A) & 110kpc\_seplim\_zA & Angular scale of 110 kpc at z\_A \\
flux\_ivar\_W2\_A & Inverse variance of W2-band flux (A) & 110kpc\_seplim\_zB & Angular scale of 110 kpc at z\_B \\
MORPHTYPE\_A & Morphological type of QSO A & delta\_vr & Velocity difference along the line of sight (km/s) \\
SPECTYPE\_A & Spectral classification of QSO A & rp\_zA & Projected distance (A--B) calculated from z\_A \\
LASTNIGHT\_A & Observation date of QSO A & rp\_zB & Projected distance (A--B) calculated from z\_B \\
TARGETID\_B & DESI target ID of QSO B & match\_type & New discovery or classification label given by previous catalogs \\
CoordUqName\_B & Coordinate name of QSO B & match\_cite & Reference source of the match label (e.g., SLED, BigMAC, NED) \\
RA\_B & Right Ascension of QSO B (deg) & visual\_class & Visual classification (QP, QPC, LQC, Contaminant) \\
DEC\_B & Declination of QSO B (deg) &  &  \\
\hline
\end{tabular}
\end{table*}

\section{Results and Discussion} \label{sec3}

\subsection{Magnitude and Angular Separation Distributions} \label{sec3.1}

\begin{figure*}
    \centering
    \includegraphics[width=\textwidth]{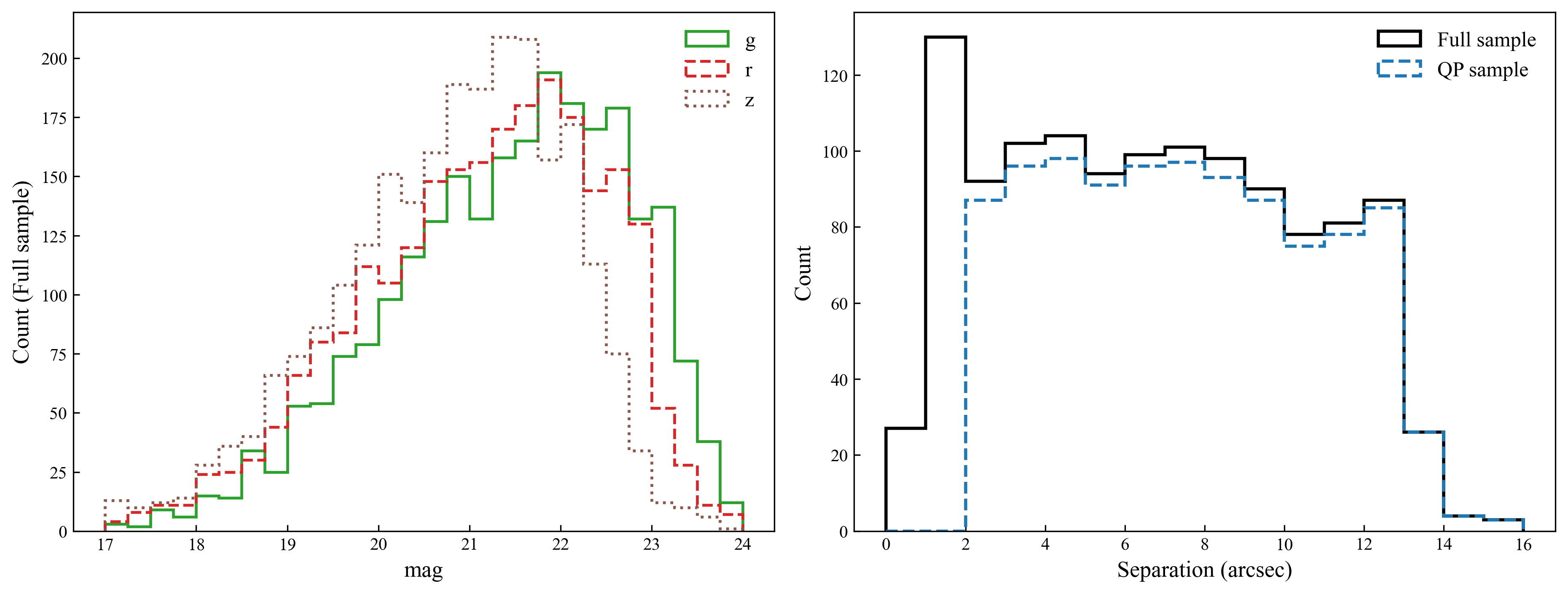}
    \caption{Left: magnitude distributions of the full sample in the $g$-, $r$-, and $z$- bands, where the magnitudes of both components of each pair are included. The sample spans $\sim$17--24 mag, with most sources clustered around $\sim21-23$ mag. Right: angular separation distributions for the full sample and the QP sample. The excess at separations below $2^{\prime\prime}$ in the full sample mainly reflects the QPC selection at very small separations, while systems at $2-13\arcsec$ are uniformly distributed.}
    \label{mag_sep_distribution}
\end{figure*}

Figure~\ref{mag_sep_distribution} gives the magnitude and angular separation distributions of the sample.
The left panel of Figure~\ref{mag_sep_distribution} shows the $g$-, $r$-, and $z$-band magnitude distributions for the full sample (QP, QPC, LQC). In constructing these histograms, we count both components of each system, so that the bins reflect the overall magnitude distribution of all quasars in the sample. Since the QP sample follows a very similar magnitude distribution and does not change the overall shape, we only show the full sample distributions for clarity. The $g$-, $r$-, and $z$-band magnitudes span $\sim 17–24$ mag, with most sources clustered around $\sim 21–23$ mag, consistent with the DESI depth and its sensitivity to faint targets \citep{2016arXiv161100036D,2025arXiv250314745D}.

The right panel of Figure~\ref{mag_sep_distribution} presents the angular separation distributions for both the full sample and the QP sample.
The full sample exhibits an excess at separations below $2^{\prime\prime}$, which is due to the QPC sample that includes systems at very small separations.
For separations of $\approx 2$–13$\arcsec$, the counts are relatively uniform.

\subsection{Transverse Distance and Velocity Difference Distributions} \label{sec3.2}

\begin{figure*}
    \centering
    \includegraphics[width=0.85\textwidth]{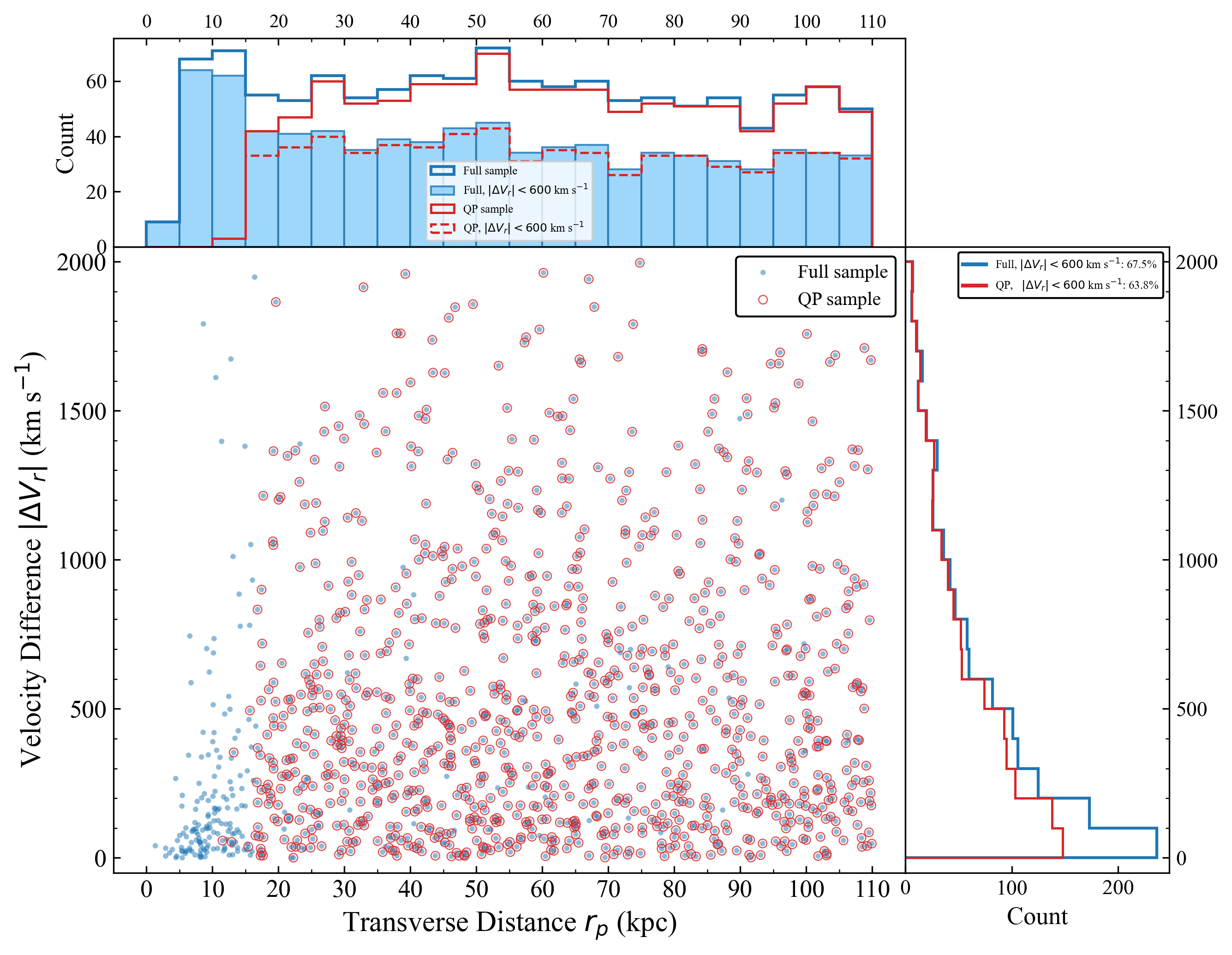}
    \caption{
    Distribution of transverse distance $r_p$ vs. velocity difference $|\Delta V_r|$ map for the full and QP samples. In the large panel, the blue points indicate the full sample, and red open circles highlight the QP sample.
    The upper histogram shows the $r_p$ distributions of the full sample (blue outline), systems with $|\Delta V_r| < 600~\mathrm{km\,s^{-1}}$ (blue filled bars), and the QP sample (solid red), with QP systems satisfying $|\Delta V_r| < 600~\mathrm{km\,s^{-1}}$ shown as a dashed red curve.
    The right histogram shows the distribution of $|\Delta V_r|$ for the full sample (blue) and the QP sample (red); the legend reports the fraction of systems with $|\Delta V_r| < 600~\mathrm{km\,s^{-1}}$ in each population.
    }
    \label{fig:rp_dv}
\end{figure*}

Figure~\ref{fig:rp_dv} shows the velocity difference $|\Delta V_r|$ vs. transverse distance $r_p$ distribution for our samples. Our samples populate a broad range in both $r_p$ and $|\Delta V_r|$, with a higher density at low $|\Delta V_r|$. Only a small number of systems occupy the high velocity--large distance or high velocity--small distance regimes, for which the bulk are not expected to be associated with ongoing mergers, in agreement with the recent results of \citet{2024MNRAS.529.1493P} and \citet{2025ApJS..281...25P}.
In terms of transverse distance, the number of quasar pairs remains roughly constant beyond $\sim 15$ kpc, and the QP sample closely follows the same $r_p$ distribution as the full sample.
The QPC systems are selected at small angular separations and therefore mostly fall at $r_p \lesssim 15$ kpc, yielding an excess of $r_p \lesssim 15$ kpc pairs in the full sample and making them promising candidates for strongly interacting quasar pairs.

For velocity difference, previous studies have suggested that the commonly adopted threshold of 2000\,km\,s$^{-1}$ for radial velocity differences \citep[e.g.,][]{2006AJ....131....1H,2010ApJ...719.1672H} may be too generous for identifying physically associated systems, and suggested a tight threshold of 600\,km\,s$^{-1}$ \citep{2011ApJ...737..101L,2025ApJS..281...25P}. Since our sample typically lies at high redshifts (see Figure~\ref{zdistribution}) and the broad emission lines are used to trace the redshifts, we adopted the threshold of 2000\,km\,s$^{-1}$ to ensure the completeness of our quasar pair selection. However, to further assess the reliability of our analysis, we also used $|\Delta V_r| < 600$\,km\,s$^{-1}$, a more stringent criterion favored by some works \citep{2011ApJ...737..101L,2024arXiv241208397J,2025ApJS..281...25P}, to select sub-samples from both the full and QP samples and to examine their transverse distance distributions. The results are overplotted in the upper histogram of Figure~\ref{fig:rp_dv}. For $r_p \gtrsim 15$\,kpc, the $|\Delta V_r| < 600$\,km\,s$^{-1}$ samples in both the full and QP samples closely follow the nearly flat $r_p$ distribution of the parent samples, indicating that our results are not biased by the choice of the velocity threshold. We also quantified the incidence of low velocity difference systems, giving that $67.5\%$ of the full sample and $63.8\%$ of the QP sample satisfy $|\Delta V_r| < 600$\,km\,s$^{-1}$. The corresponding $|\Delta V_r|$ distributions are shown in the right-hand panel of Figure~\ref{fig:rp_dv}.

Our pair counts are nearly flat at $r_p \gtrsim 15$ kpc in Figure~\ref{fig:rp_dv}, consistent with the low-redshift ($z<0.5$) dual AGNs selected from the Million Quasar Catalog by \citet{2026arXiv260104627D}. The agreement is encouraging given that our sample is selected only from DESI, and thus it suffers fewer heterogeneous selection effects. 
In practice, different gas inflow mechanisms could allow for time lags between torquing and AGN fueling and/or more variable AGN fueling \citep{2015MNRAS.451.2517S,2018MNRAS.479.3952B,2019ApJ...879L..21G,2025ApJS..281...25P}. Such cascades can smear the observable activity over a wide range of orbital phases. It is therefore plausible to detect dual AGNs not only around the first close passage but also at later stages between pericenter and apocenter and during subsequent encounters \citep{2023MNRAS.522.1895C,2025ApJS..281...25P}. The nearly flat distribution of $r_p$ suggests the quasars are being continuously fueled or their gas reserves are being replenished throughout the merger cycle, which means there are internal torques (triggered initially during the pericenter passage(s)) and likely also tidal debris, so the gas distributions and overall structures of the hosts are changing with time. Our results support the AGN excess reported at large distance ($\gtrsim100$~kpc) in Illustris--TNG100 galaxy pairs \citep{2024MNRAS.528.5864B}, and also agree with observational evidence for enhanced AGN incidence out to $\sim$40–50 kpc and $\sim$60 kpc in mid-IR \citep{2014MNRAS.441.1297S} and optical \citep{2011MNRAS.418.2043E} merger samples. 

Nevertheless, our result seems different from several studies that report an enhancement toward a smaller transverse distance for dual AGNs.
For example, \citet{2011ApJ...737..101L} constructed a sample of 1286 AGN pairs from SDSS-DR7 with a velocity difference of $|\Delta V_r|<600~\mathrm{km\,s^{-1}}$, the majority of which lie at $z<0.16$. They found that the number of AGN pairs increases toward smaller transverse distances after completeness correction. Using archival \textit{Chandra} observations of this optically selected AGN-pair sample, \citet{2020ApJ...900...79H} further reported that the 2--10\,keV X-ray luminosity increases with decreasing transverse distance for $r_p \gtrsim 15$~kpc, while the trend may reverse at $r_p \lesssim 15$~kpc. Likewise, \citet{2023MNRAS.519.5149D}, drawing on the same SDSS-selected AGN pairs, found increasing intrinsic X-ray luminosity with decreasing $r_p$, and that systems at smaller $r_p$ exhibit higher obscuration than those with $r_p \gtrsim 50$--60~kpc.
Meanwhile, \citet{2012ApJ...746L..22K} found that both the pair fraction and the mean hard X-ray luminosity increase strongly toward smaller projected separations for all-sky \textit{Swift}/BAT-selected AGNs and SDSS Seyferts at $z<0.05$.
However, compared to their low-redshift AGN pairs, our targets are distributed predominantly at much higher redshifts (most at $z \sim 1$–$2.5$; see Figure~\ref{zdistribution}), implying that the underlying sample properties are substantially different. This motivates more dedicated physical follow-up to assess how the triggering, obscuration, and fueling of SMBH accretion evolve with time in quasar pairs. It is also worth noting that our sample may be affected by potential selection bias, with incompleteness and contamination that are difficult to quantify—particularly at small angular separations where identification is strongly impacted by fiber spillover \citep{2020A&A...639A.117H,2023ApJ...945..167P} and observing conditions \citep{2019AJ....157..168D,2025arXiv250314745D}. Moreover, since our sample is selected purely from DESI optical quasars—which primarily targets unobscured, optically bright AGN \citep{2016arXiv161100036D,2025arXiv250314745D}—and we lack a dedicated multi-wavelength analysis, we may miss a substantial population of quasar pairs in which at least one component is obscured, a phase expected to be more common at small separations \citep{2019ApJ...875..117P,2023MNRAS.519.5149D}.

As a result, the distributions of $r_p$ and $|\Delta V_r|$ indicate that a substantial fraction of our sample likely consists of physically associated quasar pairs, which provide a means to study potentially merger-triggered or merger-induced SMBH growth across the merger sequence. A more detailed analysis, potentially accounting for merger time scales, observational resolution limits, and selection effects, is required to draw robust conclusions \citep{2025ApJS..281...25P}.

\subsection{Redshift evolution of the quasar pair fraction}
\label{sec3.3}

\begin{figure}[H]
    \centering
    \includegraphics[width=0.48\textwidth]{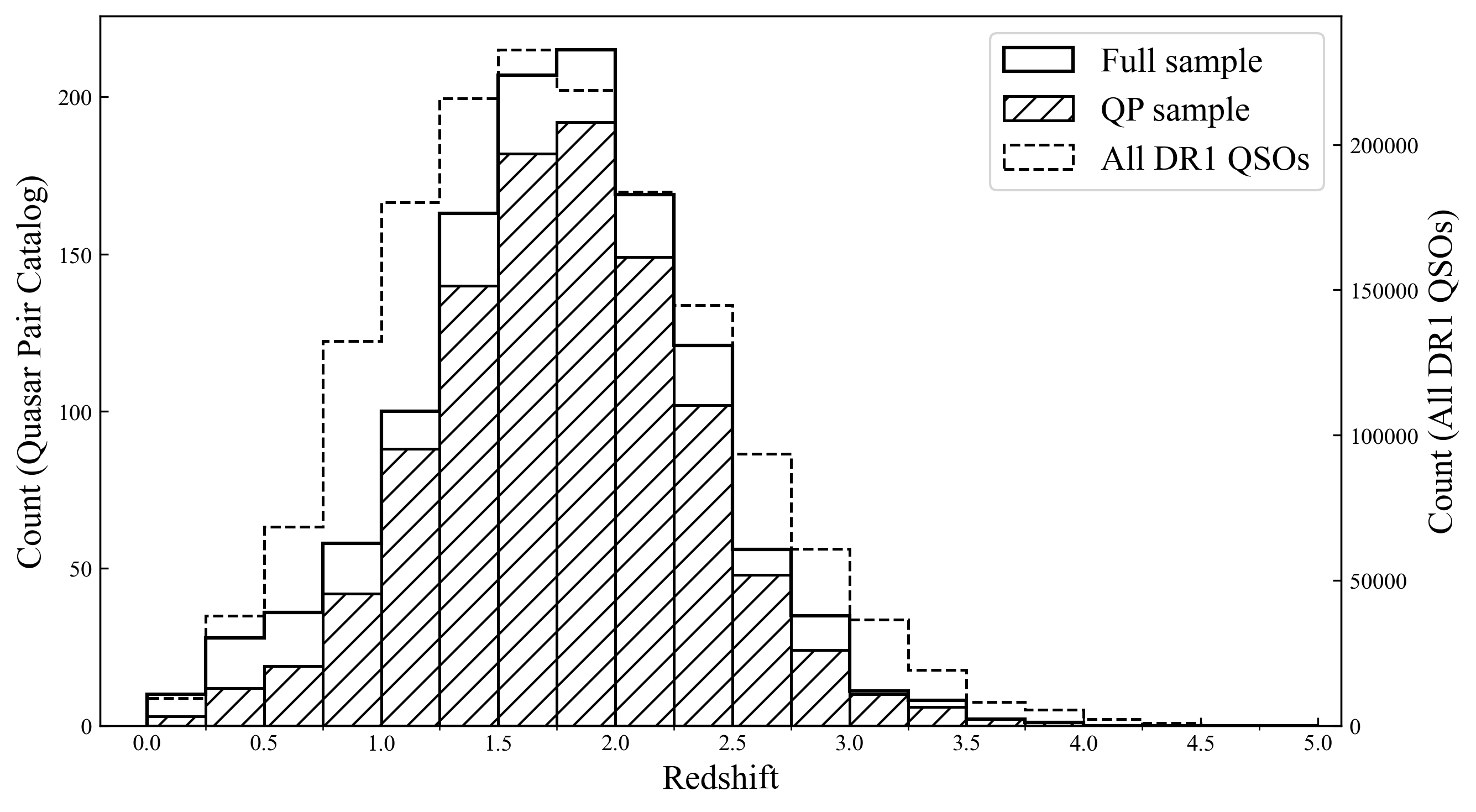}
    \caption{Redshift distribution of the full (QP, QPC, LQC) and QP samples, along with that of all spectroscopically confirmed quasars in DESI DR1 for reference. The QP sample is primarily concentrated in the redshift range $z \sim 1$–2.5. The left y-axis corresponds to the counts of the full and QP samples, while the right y-axis shows the counts of all DESI DR1 quasars, overplotted for comparison.}
    \label{zdistribution}
\end{figure}

The redshift distribution of our quasar pair sample is shown in Figure \ref{zdistribution}. The majority of sources in both the full (QP, QPC, LQC) and QP samples are found in the redshift range $z \sim 1$–2.5, coinciding with the peak epoch of quasar activity and cosmic assembly \citep{2000MNRAS.311..576K,2006AJ....131.2766R,2008ApJS..175..356H}. For reference, we overlay the redshift distribution of all DESI DR1 quasars.

\begin{figure}[H]
    \centering
    \includegraphics[width=0.48\textwidth]{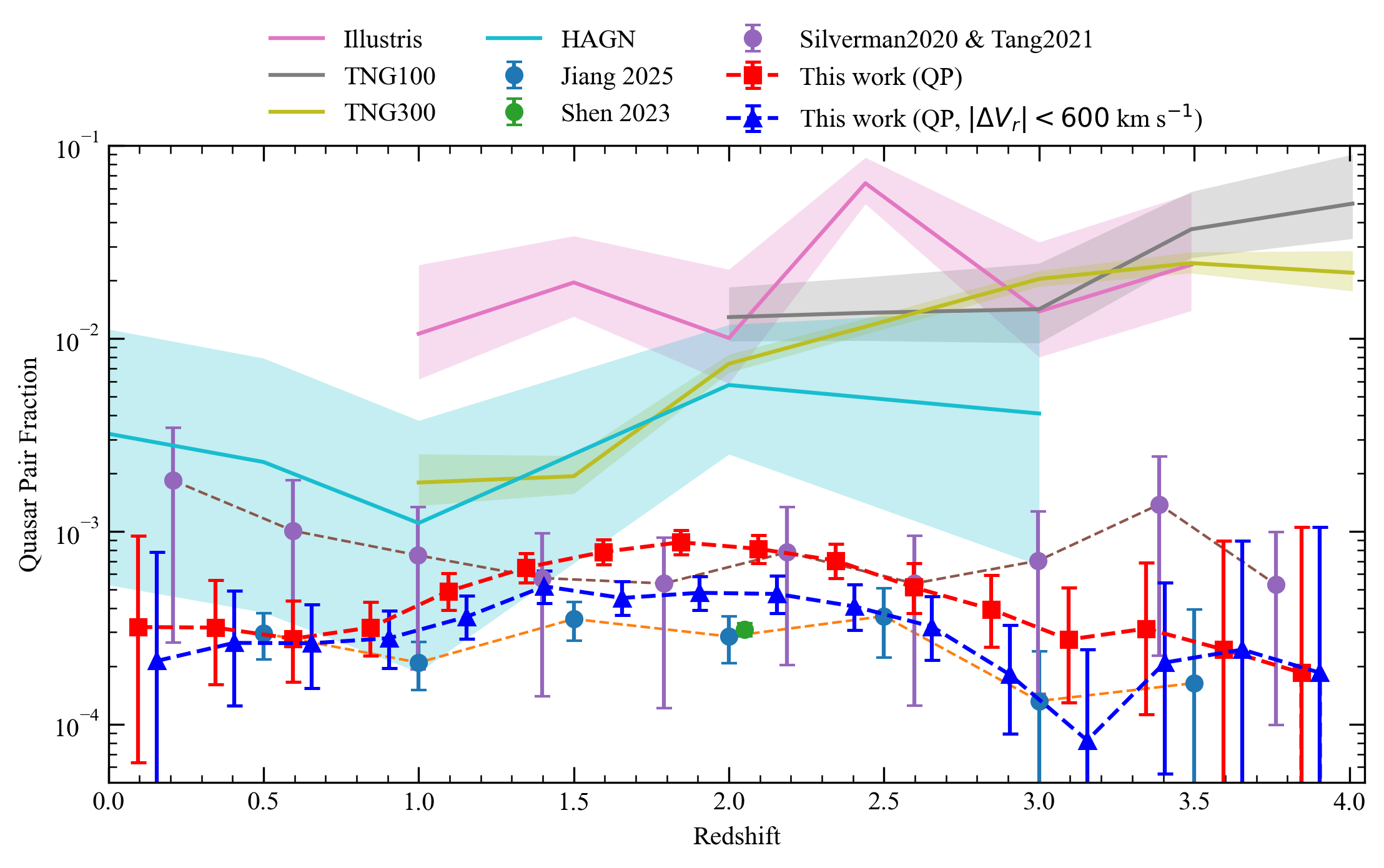}
    \caption{Quasar-pair fraction as a function of redshift.
    The red squares show the QP fraction measured in this work, while the blue triangles show the fraction for the subset with $|\Delta V_r|<600~{\rm km~s^{-1}}$. Our fractions are computed in fixed $\Delta z=0.25$ bins, and the two series are shown with a small visual redshift-offset for clarity.
    For comparison, we plot previous observational measurements from \citet{2020ApJ...899..154S} (purple circle), \citet{2023ApJ...943...38S} (green circle), and \citet{2025arXiv250406415J} (light blue circle).
    We also include simulation-based predictions matched to the same observational selection as \citet{2020ApJ...899..154S}, including Illustris (magenta), TNG100 (gray), TNG300 (olive), and HAGN (cyan), with fractions derived by \citet{2025MNRAS.536.3016P}. The error bars or shaded regions indicate Poisson uncertainties for each dataset.}
    \label{fraction}
\end{figure}

Following the works in \citet{2025MNRAS.536.3016P} and \citet{2025arXiv250406415J}, we investigated the evolution of quasar pair fraction for our sample.
For consistency with the literature, we restrict the analysis to the pure QP systems, thereby minimising contamination from lensed quasars. It is worth noting that this is only a simple, first-order calibration: the choice of pure QP systems may miss some genuine quasar pairs of QPC and LQC systems with very small separations, and may also include some sources misclassified as QP due to spectral spillover effects \citep{2020A&A...639A.117H,2023ApJ...945..167P}. We binned the sample with a redshift bin size of 0.25 and estimated the uncertainty in each bin from the 2$\sigma$ Poisson confidence interval of the pair counts. The results are shown in Figure~\ref{fraction} as red points.

For comparisons, Figure~\ref{fraction} also shows some results of previous observational studies \citep{2020ApJ...899..154S,2021ApJ...922...83T,2023ApJ...943...38S} and of large-scale cosmological simulations, including Illustris \citep{2014MNRAS.445..175G,2014MNRAS.444.1518V}, TNG100, TNG300 \citep{2018MNRAS.473.4077P}, and Horizon-AGN (HAGN; \citealp{2016MNRAS.463.3948D,2016MNRAS.460.2979V}). 
Our quasar pair fraction remains at the level of a few $\times 10^{-4}$ over each redshift bin, with an overall fraction of $6.2^{+0.2}_{-0.2}\times10^{-4}$. This result is broadly consistent, at the $1\sigma$ level, with Gaia selected double quasars by  \citet{2023ApJ...943...38S} ($z > 1.5 \ \mathrm{and}\ G < 20.25$) and \citet{2025arXiv250406415J} ($z > 0.5 \ \mathrm{and}\ G < 20.5$), as well as with the Subaru HSC imaging selected dual AGNs \citep{2020ApJ...899..154S,2021ApJ...922...83T} ($z \le 3.5 \ \mathrm{and}\ L_{\rm bol} > 10^{44.3}\ {\rm erg\ s^{-1}}$). Our slightly higher fraction than other observational studies may arise from the enhanced sensitivity to fainter quasars of DESI (see Figure \ref{mag_sep_distribution}). In contrast, large-volume simulations (Illustris/TNG/HAGN) typically predict a higher fraction than observed, likely because optical selection can miss a lot of obscured quasars \citep{2003ApJ...582L..15K,2014MNRAS.441.1297S,2017ApJ...848..126S,2017MNRAS.468.1273R,2021MNRAS.506.5935R,2017MNRAS.469.4437C,2018MNRAS.478.3056B,2019ApJ...883..167P}.

The QP fraction in our sample shows weak redshift evolution, consistent with previous observational and simulation results \citep{2025arXiv250406415J}, although it appears mildly elevated at $z \sim 1–2.5$. To obtain a more robust assessment, we repeated the measurement using a stricter subsample requiring $|\Delta V_r|<600~{\rm km\,s^{-1}}$ (\citealp{2011ApJ...737..101L,2025ApJS..281...25P}; for more detailed discussion, see Section~\ref{sec3.2}). The resulting fraction follows the weak-evolution trend found by \citet{2025arXiv250406415J} and \citet{2020ApJ...899..154S} much more closely. Taken together, the mild elevation of the QP fraction at $z\sim1–2.5$ may suggest: (1) a non-negligible contribution from high-$|\Delta V_r|$ fake systems that biases the statistics; (2) a modest enhancement at cosmic noon that could reflect heightened quasar activity and its potential role in facilitating mergers; and (3) selection bias in our optically selected sample, which preferentially targets unobscured AGN and may therefore introduce biases.
Overall, our results support a weak redshift evolution of the quasar-pair fraction, while also highlight the likely impact of selection effects, motivating a more careful investigation with larger, multi-wavelength–selected samples.

\subsection{A Wide-Separation Quadruple Lens Candidate: J1011$-$0505} \label{sec3.4}
\begin{figure}[H]
    \centering
    \includegraphics[width=0.48\textwidth]{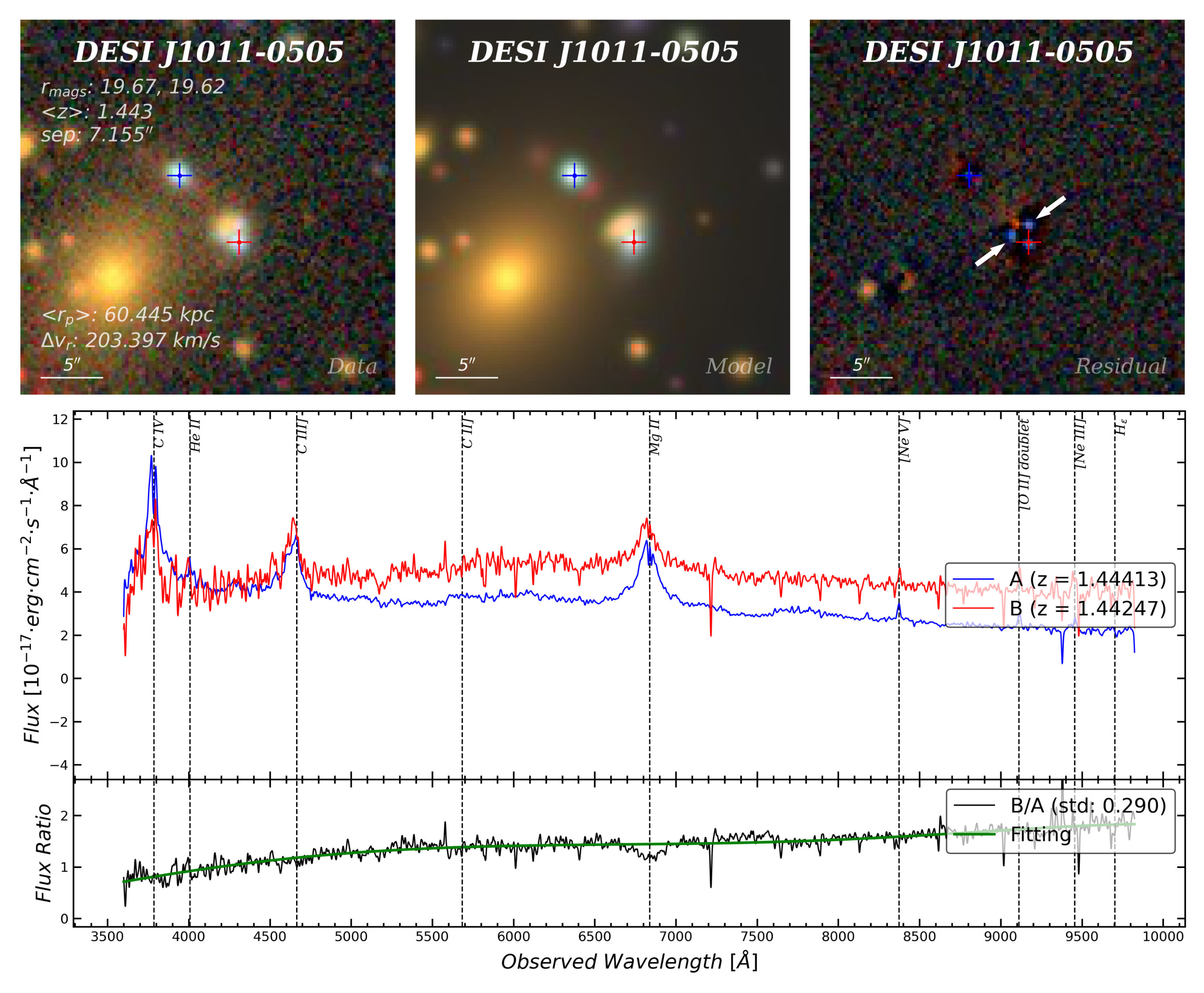}
    \caption{J1011$-$0505, a wide-separation quadruple lens candidate selected from the LQC sample in Subsection~\ref{sec2.4.3}. The image types of all panels are the same as in Figure \ref{ex_QP}. The quasars exhibit a wide separation and appear in a symmetric configuration. Notably, both spectra show a consistent self-absorption feature at the location of the \mbox{Mg \textsc{II}} emission line, potentially indicative of lensed images of the background quasar. The residual map reveals two small blue excesses near the location of source B, as marked by white arrows, suggesting the presence of additional images. Several galaxies are also observed near the sources at redshift $z \sim 0.32$, indicative of a group- or cluster-scale lensing scenario.}
    \label{J1011}
\end{figure}

J1011$-$0505, shown in Figure~\ref{J1011}, was identified during our visual classification as a particularly intriguing system: a wide-separation ($\sim$7.15$^{\prime\prime}$) four-image lens candidate in the LQC sample (Subsection \ref{sec2.4.3}). The system displays a symmetric configuration, with two components (the two crosses) exhibiting nearly identical spectral features. The residual image reveals two additional flux peaks (marked by white arrows) near quasar B, indicating a quadruply imaged configuration. 
What makes this candidate stand out is its large image separation and apparent morphology. Typical galaxy-scale quadruply lensed quasars have image separations of order $\sim$1--2$^{\prime\prime}$ and are commonly described by the standard fold/cusp or Einstein-cross geometry \citep[e.g.,][]{2024SSRv..220...23L}. In contrast, the $\sim$7.15$^{\prime\prime}$ separation of J1011$-$0505 is difficult to produce with an isolated galaxy potential and instead suggests a more massive deflector, such as a galaxy group/cluster or a multi-deflector environment that can yield wide-separation quasar images (e.g., \citealp{2003Natur.426..810I,2006ApJ...653L..97I,2013ApJ...773..146D,2018MNRAS.481L.136S,2019MNRAS.489.4741S,2021ApJ...921...42S,2023ApJ...946...63M,2023ApJ...954L..38N}). Interestingly, J1011$-$0505 exhibits an unusual quad configuration: the images form a wide, Einstein-cross morphology, yet two of the putative images are separated by only a very small distance. This is uncommon among standard quad lenses (see many examples in \citealp{2024SSRv..220...23L}), thereby enhancing its interest as a rare lensing system.
We performed a preliminary check of the system environment using the available DESI imaging and spectroscopic data. The lens resides in a dense galaxy environment consistent with a galaxy group or cluster, with DESI DR1 spectroscopic redshifts $z \sim 0.32$. We are currently undertaking follow-up analyses, including deep imaging, spectroscopic observations, and gravitational lens modeling, to verify the lensing nature.

\section{Summary} \label{sec4}

In this study, we present a large, systematically constructed sample of quasar pairs (QPs) based on the DESI DR1 quasar sample \citep{2025arXiv250314745D}, which includes 1.6 million spectroscopically confirmed quasars. Using a redshift-dependent self-matching strategy, we applied physical criteria on transverse distance ($\lesssim 110$\,kpc; \citealp{2025ApJS..281...25P}) and radial velocity difference ($\lesssim 2000$\,km\,s$^{-1}$; \citealp{2006AJ....131....1H,2010ApJ...719.1672H}) and identified 1,220 quasar pairs or candidates. This constitutes a large, homogeneous, high-redshift sample of spectroscopically selected quasar pairs and candidates from DESI DR1. To assess overlap with known systems, we searched archival records for duplicate identifications of our samples. In total, these cross-matches identify 145 systems with previous classifications, including 23 confirmed lensed quasars and 122 candidate lensed quasars or quasar pairs reported in the literature.

We conducted a comprehensive visual classification based on DESI Legacy Imaging (LS-DR9) and SPARCL spectra, dividing the sample into three categories: QP (quasar pairs, N = 1020), QPC (quasar pair candidates, N = 142), LQC (lensed quasar candidates, N = 58). For each category, we give an example system, including images and spectra. Some basic statistics on the samples include:

\begin{itemize}
    \item The sample spans $g$, $r$, and $z$ magnitudes of $\sim$17–24 mag (peaking at $\sim$21–23 mag), and the angular separation distribution is approximately flat over $\sim$2–13$\arcsec$.
    
    \item The sample span a broad redshift range up to $z<5$, with the majority found at $z\sim1-2.5$, coinciding with the peak epoch of quasar activity. Based on our DESI DR1 quasar catalog, we estimate an overall quasar pair fraction of $6.2^{+0.2}_{-0.2}\times10^{-4}$ (Poisson error) using the QP class. The pair fraction exhibits weak evolution with redshift, consistent with previous observational measurements and theoretical simulations \citep{2025MNRAS.536.3016P,2025arXiv250406415J}.

    \item The distribution of pairs shows no strong trend with $r_p$ above $\sim$15 kpc, which may suggest sustained quasar fueling or continued gas replenishment across the merger cycle. Our results support the AGN excess reported at large distance \citep{2011MNRAS.418.2043E,2014MNRAS.441.1297S,2024MNRAS.528.5864B}.

    \item Approximately 63.8\% of QP systems exhibit radial velocity differences below 600\,km\,s$^{-1}$, supporting the hypothesis that a significant fraction are physically associated and not merely projected alignments.

\end{itemize}

This sample substantially expands the census of high redshift quasar pairs, provides a large statistical basis for studying the formation and evolution of galaxies and the co-evolution of supermassive black holes and their host galaxies. Notably, the sample includes a rare wide-separation quadruply imaged lens candidate, which we are currently investigating through follow-up analysis. Future follow-up with high-resolution imaging and multi-wavelength data will further clarify the nature of these systems and constrain models of AGN triggering and merger evolution.

\section*{Data Availability}
The quasar pair sample in this work is available and can be downloaded in \url{https://github.com/astroliang/DESI_DR1_QP}

\section*{Acknowledgments}

We are very grateful to the anonymous referee for an exceptionally detailed and insightful review. The referee provided numerous constructive comments and valuable suggestions, which greatly strengthened the presentation and significantly improved the manuscript. We thank Yuanzhe Jiang for sharing figure data on dual quasar fractions.

This work has been supported by the Chinese National Natural Science Foundation grant No. 12333001 and by the National Key R\&D Program of China (2021YFA0718500 and 2025YFA1614101).
Hu Zou acknowledges the supports from National Key R\&D Program of China (grant No. 2023YFA1607804 and 2022YFA1602902), and National Natural Science Foundation of China (NSFC; grant Nos. 12120101003, and 12373010). Jun-Qing Xia acknowledges the supports from the China Manned Space Program, grant Nos. CMS-CSST-2025-A01 and CMS-CSST-2025-A04.

This research uses services or data provided by the SPectra Analysis and Retrievable Catalog Lab (SPARCL) and the Astro Data Lab, which are both part of the Community Science and Data Center (CSDC) program at NSF National Optical-Infrared Astronomy Research Laboratory. NOIRLab is operated by the Association of Universities for Research in Astronomy (AURA), Inc. under a cooperative agreement with the National Science Foundation.

The DESI Legacy Imaging Surveys consist of three individual and complementary projects: the Dark Energy Camera Legacy Survey (DECaLS), the Beijing-Arizona Sky Survey (BASS), and the Mayall z-band Legacy Survey (MzLS). Pipeline processing and analyses of the data were supported by NOIRLab and the Lawrence Berkeley National Laboratory (LBNL). Legacy Surveys was supported by: the Director, Office of Science, Office of High Energy Physics of the U.S. Department of Energy; the National Energy Research Scientific Computing Center, a DOE Office of Science User Facility; the U.S. National Science Foundation, Division of Astronomical Sciences; the National Astronomical Observatories of China, the Chinese Academy of Sciences and the Chinese National Natural Science Foundation. LBNL is managed by the Regents of the University of California under contract to the U.S. Department of Energy.

This research used data obtained with the Dark Energy Spectroscopic Instrument (DESI). DESI construction and operations is managed by the Lawrence Berkeley National Laboratory. This material is based upon work supported by the U.S. Department of Energy, Office of Science, Office of High-Energy Physics, under Contract No. DE–AC02–05CH11231, and by the National Energy Research Scientific Computing Center, a DOE Office of Science User Facility under the same contract. Additional support for DESI was provided by the U.S. National Science Foundation (NSF), Division of Astronomical Sciences under Contract No. AST-0950945 to the NSF’s National Optical-Infrared Astronomy Research Laboratory; the Science and Technology Facilities Council of the United Kingdom; the Gordon and Betty Moore Foundation; the Heising-Simons Foundation; the French Alternative Energies and Atomic Energy Commission (CEA); the National Council of Humanities, Science and Technology of Mexico (CONAHCYT); the Ministry of Science and Innovation of Spain (MICINN), and by the DESI Member Institutions: www.desi.lbl.gov/collaborating-institutions. The DESI collaboration is honored to be permitted to conduct scientific research on I’oligam Du’ag (Kitt Peak), a mountain with particular significance to the Tohono O’odham Nation. Any opinions, findings, and conclusions or recommendations expressed in this material are those of the author(s) and do not necessarily reflect the views of the U.S. National Science Foundation, the U.S. Department of Energy, or any of the listed funding agencies.

\begin{contribution}
Liang Jing and Qihang Chen contributed equally to this work and should be considered co-first authors.
Zhuojun Deng and Xingyu Zhu assisted in the visual inspection and classification of quasar pair candidates, and Xingyu Zhu also contributed to the implementation of the machine-learning pipeline.
Jun-Qing Xia provided technical support for the machine-learning implementation.
Hu Zou provided guidance on the use and interpretation of DESI imaging and spectroscopic data.
Jianghua Wu supervised all stages of the research.


\end{contribution}

%
\facility{Astro Data Lab}

\software{SPARCL \citep{2024arXiv240105576J}}

\appendix

\section{Contaminant}\label{app:contaminant}
In addition to the 1,220 quasar pairs and candidates that satisfy transverse distance $r_p \lesssim 110$ kpc \citep{2025ApJS..281...25P} and velocity difference $|\Delta V_r| \lesssim 2000$ km s$^{-1}$ \citep{2006AJ....131....1H,2010ApJ...719.1672H}, there are 622 additional samples that satisfy the same criteria; these systems are most plausibly either duplicated observations of a single source or cases with poor spectral quality, both of which preclude reliable classification. We therefore treat them as contaminants. Although a small fraction are labeled by SLED/Big MAC as quasar pair or lensed quasar candidates, such classifications are not considered robust in our sample given the close-separation duplication/blending and the low quality spectra. 
Below we give an example of a typical contaminant.

\begin{figure}[H]
    \centering
    \includegraphics[width=0.48\textwidth]{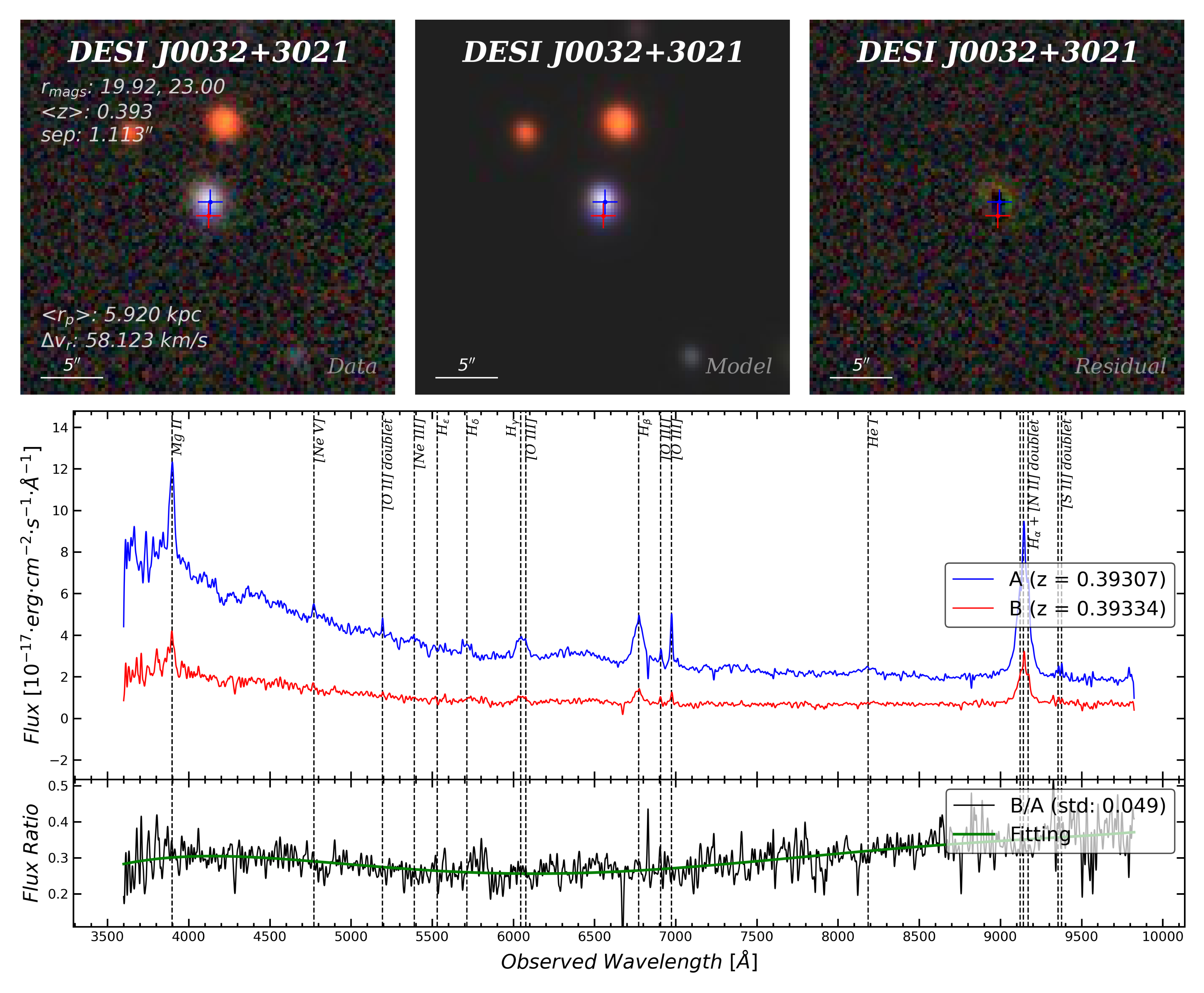}
    \caption{Example of a contaminant, J0032+3021. The two quasars have spectra with nearly identical redshifts and spectral features. The angular separation between them is only 1.11$^{\prime\prime}$, which is smaller than the DESI fiber diameter (1.5$^{\prime\prime}$). Given this small separation and the high spectral similarity, it is very plausible that the system represents a duplicate observation of the same source rather than a true quasar pair.}
    \label{ex_Uncertain}
\end{figure}

J0032+3021, shown in Figure \ref{ex_Uncertain}, exhibits two nearly identical quasar spectra at the same redshift, with an angular separation (1.11$^{\prime\prime}$) smaller than the DESI fiber diameter. The image does not clearly resolve two distinct sources, suggesting this very likely a duplicate observation rather than a true quasar pair. We therefore classify it as a contaminant.

\section{Beyond flat cuts: a machine-learning $r_p-|\Delta V_r|$ selection}
\label{sec:ml_envelope}
Motivated by both observations and simulations, the simple flat cuts in transverse distance and velocity difference can be suboptimal for isolating physically associated quasar pairs \citep{2025ApJS..281...25P}. We therefore adopted a machine-learning approach to infer an empirical $r_p$–$|\Delta V_r|$ relation and cuts for isolating quasar pairs.

We trained supervised binary classifiers on the visually classified sample and assigned each system a “pair-likeness” score, $p_{\rm pair}$, which we used for ranking and for defining a working selection threshold. Since we did not apply explicit probability calibration, $p_{\rm pair}$ was interpreted as a relative score rather than a physical probability. All thresholding and envelope analyses were confined to the core domain $r_p\le110$~kpc and $|\Delta V_r|\le2000$ kms$^{-1}$; outside this region, the scarcity of securely confirmed pairs prevents us from defining a reliable empirical boundary, and we therefore did not attempt to define or extrapolate an empirical boundary.

The labeled set comprised 1842 systems that we visually inspected in Section \ref{sec2}. We adopt an inclusive labeling scheme—treating the full sample (QP, QPC, LQC) as positives and Contaminant as negatives—to support a high completeness candidate selection strategy. After splitting the labeled sample into training, validation, and test subsets with fractions of 0.6, 0.2, and 0.2, respectively, the resulting positive/negative counts were 725/380 in the training set, 247/122 in the validation set, and 248/120 in the test set. 
Using these labeled data, we trained the classifiers on our S2 feature set, which comprises basic pair-level observables: the transverse distance $r_p$ (\texttt{rp\_used}), the velocity difference $|\Delta V_r|$ (\texttt{delta\_vr}), the angular separation (\texttt{sep\_AB}), the mean redshift (\texttt{z\_mean}), the redshift uncertainties (\texttt{ZERR\_A}, \texttt{ZERR\_B}), magnitude differences and absolute differences in the $g$, $r$, and $z$ bands (\texttt{dmag\_g}, \texttt{admag\_g}, \texttt{dmag\_r}, \texttt{admag\_r}, \texttt{dmag\_z}, \texttt{admag\_z}), color differences and absolute differences (\texttt{dcol\_gr}, \texttt{adcol\_gr}, \texttt{dcol\_rz}, \texttt{adcol\_rz}), and the redshift-difference significance metric (\texttt{S\_dz\_over\_zerr}).

\begin{figure*}
    \centering
    \includegraphics[width=0.98\textwidth]{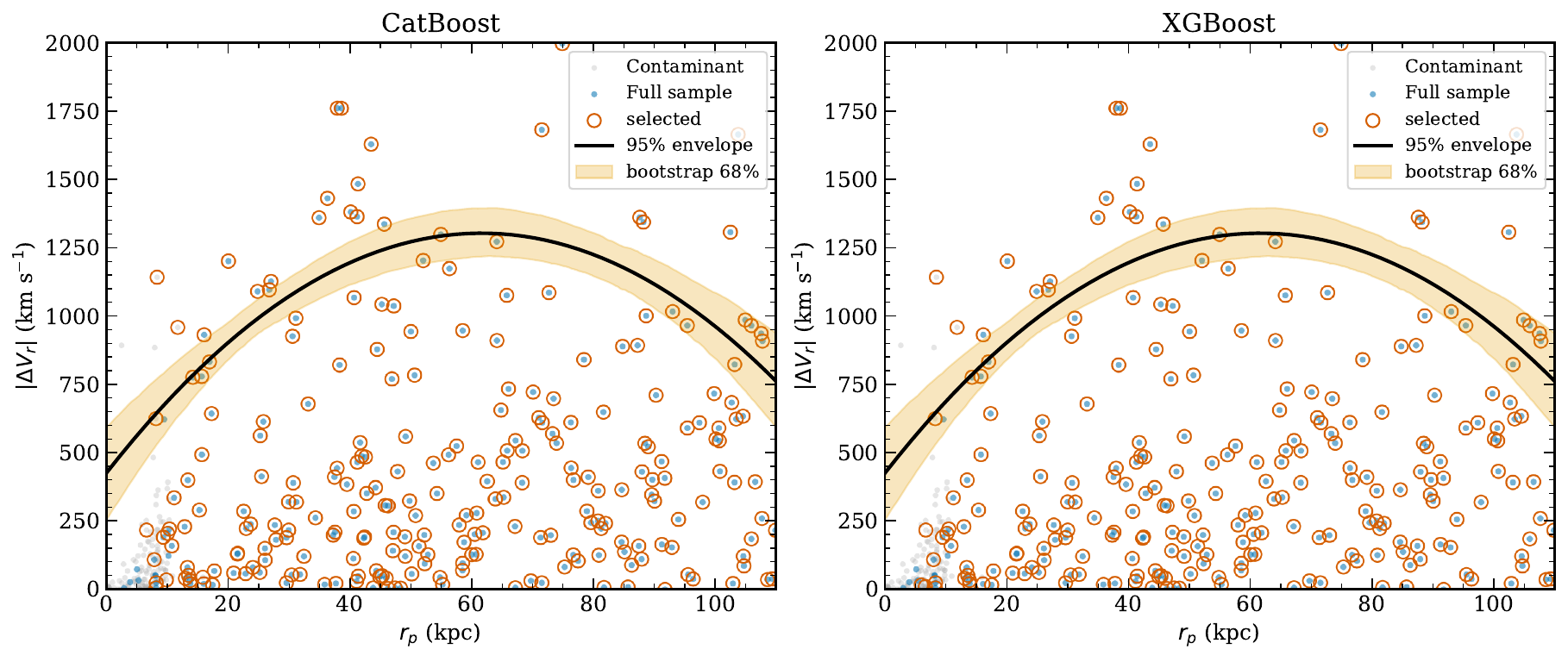}
    \caption{Distributions in the $r_p$--$|\Delta V_r|$ plane for the held-out test set and the empirical 95\% completeness envelope. The left and right panels show the results from the \textsc{CatBoost} and \textsc{XGBoost} classifiers, respectively, and they yield consistent envelopes and selections. Gray points denote test-set negatives (Contaminant), while blue points denote test-set positives (the full sample including QP, QPC, and LQC). Orange open circles mark systems selected under the fixed high-completeness threshold $t_{95}$. The black curve shows the fitted 95\% envelope, constructed by binning the recovered positives (true positives among the selected systems) in $r_p$ and taking the 95th percentile of $|\Delta V_r|$ in each bin. The shaded region shows the 68\% bootstrap uncertainty (16th–84th percentiles) from repeated binning and fitting. The envelope empirically delineates the region occupied by recovered quasar pairs at fixed $t_{95}$ within the core domain.}
    \label{rpdv_ml}
\end{figure*}

To obtain a stable feature ranking, we computed SHAP importances \citep{Lundberg2017} on the combined train$+$validation sample using GroupKFold cross-validation (implemented with \textsc{scikit-learn}; \citealt{Pedregosa2011}) and summarized the fold-wise importances (e.g., by their mean). This analysis was used only as a stability check for the adopted S2 feature set and did not involve the test set.
With the S2 feature set thus adopted, we trained \textsc{CatBoost} \citep{Dorogush2018,Prokhorenkova2018} and \textsc{XGBoost} \citep{Chen2016} classifiers (both gradient-boosted decision-tree methods; \citealt{Friedman2001}). Hyperparameters were optimized with \textsc{Optuna} \citep{Akiba2019} on the validation set (200 trials), with early stopping used to set the effective number of boosting iterations. On the held-out test set, \textsc{XGBoost} achieved $\mathrm{AUC}_{\rm ROC}=0.982$ and $\mathrm{AUC}_{\rm PR}=0.993$ \citep{Davis2006}. We then defined a high-completeness working threshold, $t_{95}$, from the combined training$+$validation predictions as the smallest score cut yielding recall $\ge 0.95$, and kept $t_{95}$ fixed on the test set for evaluation and for constructing the empirical $r_p$--$|\Delta Vr|$ envelope described below. For \textsc{XGBoost}, $t_{95}=0.190$, resulting in TP$=240$, FP$=6$, TN$=114$, and FN$=8$ (precision$=0.976$, recall$=0.968$; contamination $\simeq 2.44\%$) and selecting 246 candidates ($F_1=0.972$, MCC$=0.914$). \textsc{CatBoost} produced the same test-set confusion matrix and selected sample size, indicating that our results are insensitive to the specific boosting implementation. The corresponding test-set distributions are shown in Figure~\ref{rpdv_ml}.

To summarize the 95\% completeness boundary in the $r_p$--$|\Delta V_r|$ plane, we constructed an empirical envelope from the held-out test set. Shown in Figure \ref{rpdv_ml}, we selected the test-set positives that were recovered by the classifier under the fixed threshold $t_{95}$ (i.e., true positives among the selected systems). We then binned these recovered positives in $r_p$ and, within each bin, computed the 95th percentile of $|\Delta V_r|$. This defines a discrete upper-$|\Delta V_r|$ locus, which we denote as $|\Delta V_r|(r_p)$, i.e., the value below which 95\% of the recovered positives in that $r_p$ bin lie.
We parameterized $|\Delta V_r|(r_p)$ with a quadratic function, and the best-fitting relation is
\begin{equation}
|\Delta V_r|(r_p) \;=\; -0.23\,r_p^{2} \;+\; 28.49\,r_p \;+\; 425.66 ,
\label{eq:rpdv_quad_fit}
\end{equation}
where $r_p$ is in kpc and $|\Delta V_r|$ is in km~s$^{-1}$.
Uncertainties in Figure~\ref{rpdv_ml} were estimated via bootstrap resampling \citep{Efron1993} by repeating the binning-plus-fitting procedure and taking the 16th--84th percentiles of the resulting envelopes. Applying the fitted relation in Equation~(\ref{eq:rpdv_quad_fit}) to our raw sample, we find that 1669 out of 1842 ($\simeq 90.6\%$) systems in Section \ref{sec2.2}  satisfy $|\Delta V_r| \le |\Delta V_r|(r_p)$. This high inclusion fraction suggests that the inferred $r_p$--$|\Delta V_r|$ envelope provides an efficient, data-driven selection function that can serve as a practical alternative to simple flat cuts in transverse distance and velocity difference within the confirmed core domain.



\bibliography{sample701}{}
\bibliographystyle{aasjournalv7}


\end{document}